\documentclass[MSNbibl,nameyear,dvips]{arxstspdf}
\usepackage{flushend}
\usepackage{stfloats}
\usepackage{graphicx}

\volume{28}
\issue{4}
\pubyear{2013}
\firstpage{510}
\lastpage{520}
\doi{10.1214/13-STS440} 

\makeatletter
\def\cal{\mathcal}
\makeatother

\begin{document}
\begin{frontmatter}

\title{Assessment of Point Process Models for Earthquake Forecasting}%
\runtitle{Earthquake Model Assessment}

\begin{aug}
\author[a]{\fnms{Andrew} \snm{Bray}}
\and
\author[a]{\fnms{Frederic Paik} \snm{Schoenberg}\corref{}\ead[label=e2]{frederic@stat.ucla.edu }}
\runauthor{A. Bray and F. P. Schoenberg}

\affiliation{UCLA}

\address[a]{Andrew Bray is Ph.D. Student and Frederic Paik Schoenberg is
Professor,
UCLA,
Department of Statistics,
8125 Math Sciences Building,
Los Angeles, California 90095-1554,
USA.}
\end{aug}

%
\begin{abstract}
Models for forecasting earthquakes are currently tested pro\-spectively
in well-organized testing centers, using data collected
after the models and their parameters are completely specified.
The extent to which these models agree with the data is typically
assessed using a variety of numerical tests, which
unfortunately have low power and may be misleading for model comparison
purposes.
Promising alternatives exist, especially residual methods such as
super-thinning and Voronoi residuals.
This article reviews some of these tests and residual methods for
determining the goodness of fit of earthquake forecasting models.
\end{abstract}

%

\end{frontmatter}
%
\section{Introduction}\label{sec1}

A major goal in seismology is the ability to accurately anticipate
future earthquakes before they occur (\cite{Bol03}).
Anticipating major earthquakes is especially important, not only for
short-term response such as preparation of emergency personnel and
disaster relief, but also for longer-term preparation in the form of
building codes, urban planning and earthquake insurance (\cite{JorJon10}).
In seismology, the phrase \emph{earthquake prediction}
has a specific definition: it is the identification of a meaningfully
small geographic region and time window in which a major earthquake
will occur with very high probability. An example of earthquake
predictions are those generated by the M8 method (\cite*{KeiKos}), which issues an alarm whenever there is a suitably large
increase in the background seismicity of a region. Such alarms could
potentially be very valuable for short-term disaster preparedness, but
unfortunately examples of M8-type alarms, including the notable Reverse
Tracing of Precursors (RTP) algorithm, have generally exhibited low
reliability when tested prospectively, typically failing to outperform
naive methods based simply on smoothed historical seismicity (\cite*{Geletal97}; \cite*{ZecJor08}).

\emph{Earthquake prediction} can be contrasted with the related \emph
{earthquake forecasting}, which means the assignment of probabilities
of earthquakes occurring in broader space--time-magnitude regions. The
temporal scale of an earthquake forecast is more on par with climate
forecasts and may be over intervals that range from decades to
centuries (\cite{Hou10}).

Many models have been proposed for forecasting earthquakes, and since
different models often result in very different forecasts, the question
of how to assess which models seem most consistent with observed
seismicity becomes increasingly important.
Concerns with retrospective analyses, especially regarding data
selection, overfitting and lack of reproducibility, have motivated
seismologists recently to focus on prospective assessments of
forecasting models.
This has led to the development of the Regional Earthquake Likelihood
Models (RELM) and Collaborative Study of Earthquake Predictability
(CSEP) testing centers, which are designed to evaluate and compare the
goodness of fit of various earthquake forecasting models.
This paper surveys methods for assessing the models in these RELM and
CSEP experiments, including methods currently used by\break  RELM and CSEP and
some others not yet in use but which seem promising.

\section{A Framework for Prospective Testing}\label{sec2}

The current paradigm for building and testing earthquake models emerged
from the working group for the development of Regional Earthquake
Likelihood Models (RELM) in 2001.
As described in \citet{Fie07}, the participants were encouraged to
submit differing models, in the hopes that the competition between
models would prove more useful than trying to build a single consensus model.
The competition took place within the framework of a prospective test
of their seismicity forecasts.
Working from a standardized data set of historical seismicity,
scientists fit their models and submit to RELM a forecast of the number
of events expected within each of many pre-specified
spatial--temporal-magnitude bins.
The first predictive experiment required models to forecast seismicity
in California between 2006 to 2011 using only data from before 2006.

This paradigm has many benefits from a statistical perspective.
The prospective nature of the experiments effectively eliminates
concerns about overfitting. Furthermore, the standardized nature of the
data and forecasts facilitates the comparison among different models.
RELM has since expanded into the Collaborative Study of Earthquake
Predictability (CSEP), a global-scale project to coordinate mod\-el
development and conduct prospective testing according to community
standards (\cite{Jor06}).
CSEP serves as an independent entity that provides standardized
seismicity data, inventories proposed models and publishes the
standards by which the models will be assessed.

\section{Some Examples of Models for Earthquake Occurrences}\label{sec3}

The first predictive experiment coordinated\break  through RELM considered
time-independent spatial point process models, which can be specified
by their Papangelou intensity $\lambda(s)$, a function of spatial
location $s$. A representative example is the model specified by
Helmstetter, Kagan and
Jackson (\citeyear{HelKagJac07}) that is based on smoothing previous
seismicity. The intensity function is estimated with an isotropic
adaptive kernel
\[
\lambda(s) = \sum_{i=1}^{N}K_{d}
( s - s_i ),
\]
where $N$ is the total number of observed points, and $K_{d}$
is a power-law kernel
\[
K_{d}(s - s_i) = \frac{C(d)}{ (|s - s_i|^2 + d^2  )^{1.5}},
\]
where $d$ is the smoothing distance, $C(d)$ is a normalizing
factor so that the integral of $K_{d}(\cdot)$ over an infinite area equals
1, and $| \cdot|$ is the Euclidean norm. The estimated number of
points within the pre-specified grid cells is obtained by integrating
$\lambda(s)$ over each cell.

Models of earthquake occurrence that consider it to be a time-dependent
process are commonly variants of the epidemic-type aftershock sequence
(ETAS) model of Ogata (\citeyear{Oga88,Oga98}) (see, e.g., \cite*{HelSor03}; Ogata, Jones and Toda, \citeyear{OgaJonTod03};
\cite*{Sor05}; \cite*{VerZhu08}; Console, Murru and
Falcone, \citeyear{ConMurFal10}; \cite*{Chuetal11}; Wang, Jackson and Kagan, \citeyear{WanJacKag11}; \cite*{Weretal11}; \cite*{Zhu11}; \cite*{TiaShc12}).
According to the ETAS model, earthquakes cause aftershocks, which in
turn cause more aftershocks, and so on.
ETAS is a point process model specified by its conditional intensity,
$\lambda(s,t)$, which represents the infinitesimal expected rate at
which events are expected to occur around time $t$ and location $s$,
given the history $H_t$ of the process up to time $t$.
ETAS is a special case of the linear, self-exciting Hawkes' point
process (\cite{Haw71}), where the conditional intensity is of the form
\[
\label{hawkes} \lambda(s,t | H_t) = \mu(s,t) + \sum
_{t_i < t} g(s-s_i, t- t_i;
M_i),
\]
where $\mu(s,t)$ is the mean rate of a Poisson-distributed background
process that may in general vary with time and space, $g$ is a \textit{triggering function} which indicates how previous occurrences
contribute, depending on their spatial and temporal distances and
marks, to the conditional intensity $\lambda$ at the location and time
of interest, and $(s_i,t_i,M_i)$ are the origin times, epicentral
locations and moment magnitudes of observed earthquakes.

\citet{Oga98} proposed various forms for the triggering function, $g$,
such as the following:
\begin{eqnarray*}
g(s,t,M) = K (t+c)^{-p} e^{a(M-M_0)} \bigl(|s|^2 + d
\bigr)^{-q},
\end{eqnarray*}
where $M_0$ is the lower magnitude cutoff for the observed catalog.

The parameters in ETAS models and other spatial--temporal point process
models may be estimated by maximizing the log-likelihood,
\begin{eqnarray*}
\sum_{i=1}^{n} \log\bigl\{
\lambda(s_i, t_i)\bigr\} - \int_S
\int\lambda(s, t)\, \mathrm{d}s\, \mathrm{d}t.
\end{eqnarray*}

The maximum likelihood estimator (MLE) of a point process is, under
quite general conditions,\break  asymptotically unbiased, consistent,
asymptotically normal and asymptotically efficient (\cite{Oga78}).
Finding the parameter vector that maximizes the log-likelihood can be
achieved using any of the various standard optimization routines, such
as the quasi-Newton methods implemented in the function\break  optim($\cdot$) in \textit{R}.
The spatial background rate $\mu$ in the ETAS model can be
estimated in various ways, such as via kernel smoothing seismicity from
prior to the observation window or kernel smoothing the largest events
in the catalog, as in \citet{Oga98} or \citet{Sch03}.
Note that the integral term in the loglikelihood function can be
cumbersome to estimate, and an approximation method recommended in
\citet{Sch13} can be used to accelerate computation of the MLE.

There are of course many other earthquake forecasting models quite
distinct from the two point process models above.
Perhaps most important among these are the Uniform California
Earthquake Rupture Forecast (UCERF) models,
which are consulted when setting insurance rates and crafting building
codes (\cite{Fieetal09}). They are constructed by soliciting \textit{expert opinion}
from leading seismologists on which components should
enter the model, how they should be weighted, and how they should
interact (\cite{MarZec11}). Examples of the components
include slip rate, geodetic strain rates and paleoseismic data. Note
that some seismologists have argued that evaluating some earthquake
forecasting models such as UCERF using model validation experiments
such as RELM and CSEP may be inappropriate, though such a conclusion
seems to run counter to basic statistical and scientific principles.

Although the UCERF models draw upon diverse information related to the
geophysics of earthquake etiology, commonly used models such as ETAS
and its variants rely solely on previous seismicity for forecasting
future events.
Many attempts have been made to include covariates, but when assessed
rigorously, most predictors other than the locations and times of
previous earthquakes have been shown not to offer any noticeable
improvement in forecasting.
Recent examples of such covariates include electromagnetic signals
(\cite{Jac96}; \cite{Kag97}), radon (\cite{HauGod81}) and
water levels (\cite*{Baketal05}; \cite*{ManWan07}).
A promising exception is moment tensor information, which is now
routinely recorded with each earthquake and seems to give potentially
useful information regarding the directionality of the release of
stress in each earthquake.
However, this information appears not to be explicitly used presently
in models in the CSEP or RELM forecasts.

\section{Numerical Tests}\label{sec4}

Several numerical tests were initially proposed to serve as the metrics
by which RELM models would be evaluated (\cite{Schetal07}).
For these numerical tests, each model consists of the estimated number
of earthquakes in each of the spatial--tempo\-ral-magnitude bins, where
the number of events in each bin is assumed to follow a Poisson
distribution with an intensity parameter equivalent to the forecasted rate.

The $L$-test (or Likelihood test) evaluates the probability of the
observed data under the proposed mod\-el.
The numbers of observed earthquakes in each spatial--temporal-magnitude
bin are treated as independent random variables,
so the joint probability is calculated simply as the product of their
corresponding Poisson probabilities.
This observed joint probability is then considered with respect to the
distribution of joint probabilities generated by simulating many
synthetic data sets from the model.
If the observed probability is unusually low in the context of this
distribution, the data are considered inconsistent with the model.

The $N$-test (Number) ignores the spatial and magnitude component and
focuses on the total number of earthquakes summed across all bins.
If the proposed model provides estimates $\hat\lambda_i$ for $i$
corresponding to each of $B$ bins,
then according to this model, the total number of observed earthquakes
should be Poisson distributed with mean ($\sum_{i = 1}^B \hat\lambda_i$).
If the number of observed earthquakes is unusually large or small
relative to this distribution, the data are considered inconsistent
with the model.

The $L$-test is considered more comprehensive in that it evaluates the
forecast in terms of magnitude, spatial location and number of events,
while the $N$-test restricts its attention to the number of events. Two
additional data consistency tests were proposed to assess the magnitude
and spatial components of the forecasts, respectively: the $M$-test and
the $S$-test (Zechar, Gerstenberger and
Rhoades, \citeyear{ZecGerRho10}).
The $M$-test (Magnitude) isolates the forecasted magnitude distribution
by counting the observed number of events in each magnitude bin without
regard to their temporal or spatial locations,
standardized so that the observed and expected total number of events
under the model agree, and computing the joint (Poisson) likelihood of
the observed numbers of events in each magnitude bin.
As with the $L$-test, the distribution of this statistic under the
forecast is generated via simulation.

The $S$-test (Spatial) follows the same inferential procedure but
isolates the forecasted spatial distribution by
summing the numbers of observed events over all times and over all
magnitude ranges.
These counts within each of the spatial bins are again standardized so
that the observed and expected total number of events under the model agree,
and then one computes the joint (Poisson) likelihood of the observed
numbers of events in the spatial bins.


The above tests measure the degree to which the observations agree with
a particular model,
in terms of the probability of these observations under the given model.
As noted in \citet{Zecetal}, tests such as the $L$-test and $N$-test
are really tests of the consistency between the data and a particular model,
and are not ideal for comparing two models.
\citet{Schetal07} proposed an additional test to allow for the
direct comparison of the performance of two models: the Ratio test
($R$-test). For a comparison of models A and~B, and given the numbers of
observed events in each bin, the test statistic $R$ is defined as the
log-likelihood of the data according to model A minus the corresponding
log-likelihood for model B. Under the null hypothesis that model A is
correct, the distribution of the test statistic is constructed by
simulating from model A and calculating $R$ for each realization. The
resulting test is one-sided and is supplemented with the corresponding
test using model B as the null hypothesis.
The $T$-test and $W$-test of \citet{Rhoetal11} are very similar to the
$R$-test, except that instead of using simulations to find the null
distribution of the difference between log-likelihoods, with the $T$-test
and $W$-test, the differences between log-likelihoods within each
space--time-magnitude bin for models A and B are treated as independent
normal or symmetric random variables, respectively, and a $t$-test or
Wilcoxon signed rank test, respectively, is performed.


Unfortunately, when used to compare various models, such
likelihood-based tests suffer from the problem of variable null
hypotheses and can lead to highly misleading and even seemingly
contradictory results.
For instance, suppose model A has a higher likelihood than model B.
It is nevertheless quite possible for model A to be rejected according
to the $L$-test and model B not to be rejected using the $L$-test.
Similarly, the $R$-test with model A as the null might indicate that
model A performs statistically significantly better than model B, while
the $R$-test with model B as the null hypothesis may indicate that the
difference in likelihoods is not statistically significant.
Seemingly paradoxical results like these occur frequently, and at a
recent meeting of the Seismological Society of America, much confusion
was expressed over such results; even some seismologists quite well
versed in statistics referred to results in such circumstances as
``somewhat mixed,'' even though model A clearly fit better according to
the likelihood criterion than model B.

The explanation for such results is that the null hypotheses of the two
tests are different: when mod\-el~A is tested using the $L$-test, the null
hypothesis is model~A, and when model B is tested, the null hypothesis
is model B.
The test statistic may have very different distributions under these
different hypotheses.

Unfortunately, these types of discrepancies seem to occur frequently
and hence, the results of these numerical tests may not only be
uninformative for model comparison, but in fact highly misleading.
A~striking example is given in Figure~4 of \citet{Zecetal}, where
the Shen, Jackson and Kagan (\citeyear{SheJacKag07}) model produces the highest likelihood
of the five models considered in this portion of the analysis, and yet
under the $L$-test has the lowest corresponding $p$-value of the five models.

\section{Functional Summaries}\label{sec5}

Functional summaries, that is, those producing a function of one
variable, such as the weighted $K$-function and error diagrams,
can also be useful measures of goodness of fit.
However, such summaries typically provide little more information than
numerical tests in terms of indicating where and when the
model and the data fail to agree or how a model may be improved.

The weighted $K$-function is a generalized version of the $K$-function
of \citet{Rip76}, which has been widely used to detect clustering or
inhibition for spatial point processes. The ordinary $K$ function,
$K(h)$, counts, for each $h$, the total number of observed pairs of
points within distance $h$ of one another, per observed point,
standardized by dividing by the estimated overall mean rate of the
process, and the result is compared to what would be expected for a
homogeneous Poisson process.
The weighted version, $K_w(h)$, was introduced for the inhomogeneous
spatial point process case by Baddeley, M{\o}ller and
Waagepetersen (\citeyear{BadMllWaa00}), and is defined
similarly to $K(h)$, except that each pair of points $(s_i, s_j)$ is
weighted by $1 / [\hat\lambda(s_i) \hat\lambda(s_j)]$, the inverse of
the product of the modeled unconditional intensities at the points
$s_i$ and $s_j$.
This was extended to spatial--temporal point processes by \citet{VeeSch06}
and \citet{AdeSch09}.

Whereas the null hypothesis for the ordinary $K$-function is a
homogeneous Poisson process, in the case of $K_w$, the weighting allows
one to assess\break  whether the degree of clustering or inhibition in the
observations is consistent with what would be expected under the null
hypothesis corresponding to the model for $\hat\lambda$.
While weighted $K$-functions may be useful for indicating whether the
degree of clustering in the model agrees with that in the observations,
such summaries
unfortunately do not appear to be useful for comparisons between
multiple competing models, nor do they accurately indicate
in which spatial--temporal-magnitude regions there may be particular
inconsistencies between a model and the observations.

\begin{figure}

\includegraphics{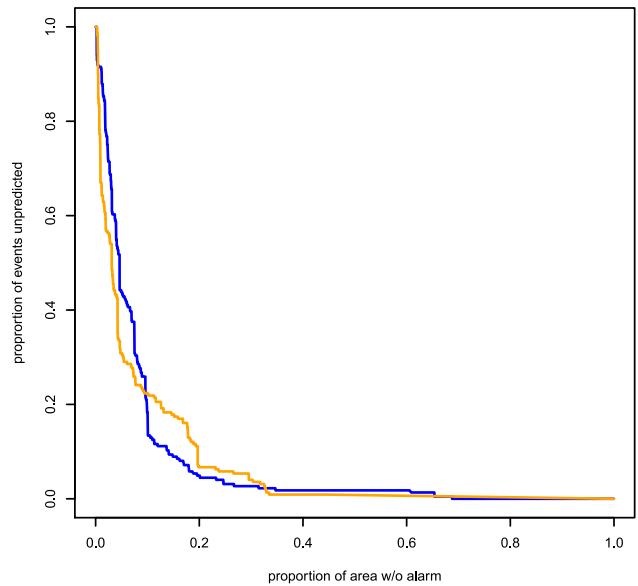}

\caption{Error diagrams for Helmstetter, Kagan and
Jackson (\citeyear{HelKagJac07}) in blue and Shen, Jackson and Kagan (\citeyear{SheJacKag07}) in orange.
Model details are in Sections~\protect\ref{sec3} and
\protect\ref{sec7}, respectively.}
\label{ED}
\end{figure}

Error diagrams, which are also sometimes called receiver operating
characteristic (ROC) curves\break  (\cite*{S73}) or Molchan diagrams (Molchan\break  (\citeyear{M91}), \citeyear{M10};
\cite*{ZM04}; \cite*{K09}),
plot the (normalized) number of alarms versus the (normalized) number
of false negatives (failures to predict), for each possible alarm,
where in the case of earthquake forecasting models an \textit{alarm}
is defined as any value of the modeled conditional rate, $\hat\lambda
$, exceeding some threshold. Figure~\ref{ED} presents error diagrams
for two RELM models, Helmstetter, Kagan and
Jackson (\citeyear{HelKagJac07}) and Shen, Jackson and Kagan (\citeyear{SheJacKag07})
(see Sections~\ref{sec3} and \ref{sec7} for model details).

The ease of interpretation of such diagrams is an attractive feature,
and plotting error diagrams with multiple models on the same plot can
be a useful way to compare the models' overall forecasting efficacy. In
Figure~\ref{ED} we learn that \citet{SheJacKag07} slightly outperforms
\citet{HelKagJac07} when the threshold for the alarm is high, but as the
threshold is lowered \citet{HelKagJac07} performs noticeably better.
For the purpose of comparing models, one may even consider normalizing
the error diagram so that the false negative rates are considered
relative to one of the given models in consideration as in \citet{K09}.
This tends to alleviate a common problem with error diagrams as applied
to earthquake forecasts, which is that most of the relevant focus is
typically very near the axes and thus it can be difficult to inspect
differences between the models graphically.
A more fundamental problem with error diagrams, however, is that while
they can be useful overall summaries of goodness of fit, such diagrams
unfortunately provide little information as to where models are fitting
poorly or how they may be improved.

\section{Residual Methods}\label{sec6}

Residual analysis methods for spatial--temporal point process models
produce graphical displays which may highlight where one model
outperforms another or where a particular model does not ideally agree
with the data.
Some residual methods, such as thinning, rescaling and superposition,
involve transforming the point process using a model for the
conditional intensity $\lambda$ and then inspecting the uniformity of
the result, thus reducing the difficult problem of evaluating the
agreement between a possibly complex spatial--temporal point process
model and data
to the simpler matter of assessing the homogeneity of the residual
point process.
Often, departures from homogeneity in the residual process can be
inspected by eye, and
many standard tests are also available.
Other residual methods, such as pixel residuals, Voronoi residuals and
deviance residuals,
result in graphical displays that can quite directly indicate locations
where a model appears to depart from the observations
or where one model appears to outperform another in terms of agreement
with the data.

\subsection{Thinned, Superposed and Super-Thinned Residuals}\label{sec6.1}

Thinned residuals are based on the technique of random thinning, which
was first introduced by\break  \citet{LewShe79} and \citet{Oga81} for the purpose
of simulating spatial--temporal point processes and extended for the
purpose of model evaluation in \citet{Sch03}.
The method involves keeping each observed point\vspace*{1pt} (earthquake)
independently with probability $b/\hat{\lambda}(s_{i}, t_{i})$,
where $b=\break  \inf_{(s, t) \in\cal S} \{\hat\lambda(s,t)\}$ and
$\hat\lambda$ is the modeled conditional intensity.
If the model is correct, that is, if the estimate $\hat{\lambda
}(s,t)=\lambda(s,t)$ almost everywhere,
then the residual process will be homogeneous Poisson with rate $b$
(\cite{Sch03}).
Because the thinning is random, each thinning is distinct, and one may
inspect several realizations of thinned residuals and analyze the
entire collection to get an overall assessment of goodness of fit, as
in \citet{Sch03}.

An antithetical approach was proposed by Bremaud (\citeyear{Bre81}), who suggested
superposing a simulated point process onto an observed point process
realization so as to yield a homogeneous Poisson process.
As indicated in \citet{CSV12}, tests based on thinned or
superposed residuals tend to have low power when the model $\hat\lambda
$ for the conditional intensity is volatile, which is typically the
case with earthquake forecasts since earthquakes tend to be clustered
in particular spatial--temporal regions. Thinning a point process will
lead to very few points remaining if the infimum of $\hat\lambda$ over
the observed space is small (\cite{Sch03}), while in superposition,
the simulated points, which are by construction approximately
homogeneous, will form the vast majority of residual points if the
supremum of $\hat\lambda$ is large.

A hybrid approach called super-thinning was introduced in \citet{CSV12}.
With super-thinning, a tuning parameter $k$ is chosen, and one thins
(deletes) the observed points in locations of space--time where $\hat
{\lambda} > k$, keeping each point independently with probability $k /
\hat\lambda(s,t)$, and superposes a Poisson process with rate $\hat
\lambda(s,t) / k$\break  where $\hat\lambda< k$.
When the tuning parameter $k$ is chosen wisely, the method appears to
be more powerful than thinning or superposing in isolation.

\subsection{Rescaled Residuals}\label{sec6.2}

An alternative method for residual analysis is re\-scaling.
The idea behind rescaled residuals dates back to \citet{Mey71}, who
investigated rescaling temporal point processes according to their
conditional intensities, moving each point $t_i$ to a new time
$\int_0 ^{t_i} \hat\lambda(t)   \,\mathrm{d} t$, creating a
transformed space in which the rescaled points are homogeneous Poisson
of unit rate.
Heuristically, the space is essentially compressed when $\hat\lambda$
is small and stretched when $\hat\lambda$ is large, so that the
points are ultimately uniformly distributed in the resulting
transformed space, if the model for $\hat\lambda$ is correct.
This method was used in \citet{Oga88} to assess a temporal ETAS model
and extended in \citet{MerNua86}, \citet{Nai90}, \citet{Sch99} and \citet{VerSch04} to the spatial and
spatial--temporal cases.
Rescaling may result in a transformed space that is difficult to
inspect if $\hat\lambda$ varies widely over the observation region,
and in such cases standard tests of homogeneity such as Ripley's
$K$-function may be dominated by boundary effects, as illustrated in
\citet{Sch03}.

\subsection{Pixel Residuals}\label{sec6.3}

A different type of residual analysis which is more closely analogous
to standard residual methods in regression or spatial statistics is to
consider the (standardized) differences between the observed and
expected numbers of points in each of various spatial or
spatial--temporal pixels or grids, producing what might be called \textit{pixel residuals}.
These types of residuals were described in great detail by \citet{Badetal05} and Baddeley, M{\o}ller and
Pakes (\citeyear{BadMllPak08}).
More precisely, the \textit{raw} pixel residual on each pixel $A_i$ is
defined as $N(A_i) - \int\hat{\lambda}(s,t)\, \mathrm{d}t\,   \mathrm
{d}s$, where $N(A_i)$ is simply the number of points (earthquakes)
observed in pixel $A_i$ (\cite{Badetal05}). \citet{Badetal05} also proposed various standardizations including Pearson
residuals, which are scaled in relation to the standard deviation of
the raw residuals:
$r_i = \frac{N(A_i) - \int\hat{\lambda}(s,t)  \, \mathrm{d}t \,  \mathrm
{d}s}{\sqrt{\int\hat{\lambda}(s,t)   \,\mathrm{d}t \, \mathrm{d}s}}$.

A problem expressed in \citet{Wonetal} is that if the pixels are
too large, then the method is not powerful to detect local
inconsistencies between the model and data, and places in the interior
of a pixel where the model overestimates seismicity may cancel out with
places where the model underestimates seismicity.
On the other hand, if the pixels are small, then the majority of the
raw residuals are close to zero while those few that correspond to
pixels with an earthquake are close to one. In these situations where
the residuals have a highly skewed distribution, the skew is only
intensified by the standardization to Pearson residuals. As a result,
plots of both the raw and the Pearson residuals are not informative and
merely highlight the pixels where earthquakes occur regardless of the
fit of the model.
The raw or Pearson residuals may be smoothed, as in \citet{Badetal05}, but such smoothing typically only reveals gross, large-scale
inconsistencies between the model and data.

If one is primarily interested in comparing competing models, then
instead one may plot, in each pixel, the difference between
log-likelihoods for the two models, as in Clements, Schoenberg and
Schorlemmer (\citeyear{CleSchSch11}). The
resulting residuals may be called \textit{deviance residuals}, in analogy
with residuals from logistic regression and other generalized linear models.
Deviance residuals appear to be useful for comparing models on grid
cells and inspecting where one model appears to fit the observed
earthquakes better than the other. It remains unclear how these
residuals may be used or extended to enable comparisons of more than
two competing models, other than by comparing two at a time.

\subsection{Voronoi Residuals}\label{sec6.4}

One method of addressing the problem of pixel size specification is to
use a data-driven, spatially adaptive partition such as the Voronoi
tessellation, as suggested in \citet{Wonetal}.
Given $n$ observed earthquakes, one may obtain a collection of $n$
Voronoi cells $A_1, \ldots, A_n$, where
$A_i$ is defined as the collection of spatial--temporal locations closer
to the particular point (earthquake) $i$ than to any of the other observed
points
(\cite{Okaetal00}).
Thus, $N(A_i) = 1$ for each cell $A_i$.
One may then compute the corresponding standardized residuals
$r_i = \frac{1 - \int\hat{\lambda}(s,t)  \, \mathrm{d}t\,   \mathrm
{d}s}{\sqrt{\int\hat{\lambda}(s,t)   \,\mathrm{d}t \,  \mathrm{d}s}}$
over the Voronoi cells $A_i$.
As with pixel residuals, for each Voronoi cell one may choose to plot
the raw residual, or the residual deviance if one is interested in
comparing competing models.
Voronoi residuals are shown in \citet{Wonetal} to be generally less
skewed than pixel residuals and are approximately Gamma distributed
under quite general regularity conditions.

\section{Examples}\label{sec7}

In the present section we apply some of the residual methods discussed
above to models and seismicity data from the 5-year RELM prediction
experiment that ran from 2006 to 2011. The original experiment called
for modelers to estimate the number of earthquakes above magnitude 4.95
that would occur in many pre-specified spatial bins in California.
During this time period only 23 earthquakes that fit these criteria
were recorded, a~fairly small data set from which to assess a model. In
order to better demonstrate the methods available in residual analysis,
the models that we consider were recalibrated using their specified
magnitude distributions to forecast earthquakes of greater than
magnitude 4.0, of which there are 232 on record.

\begin{figure*}

\includegraphics{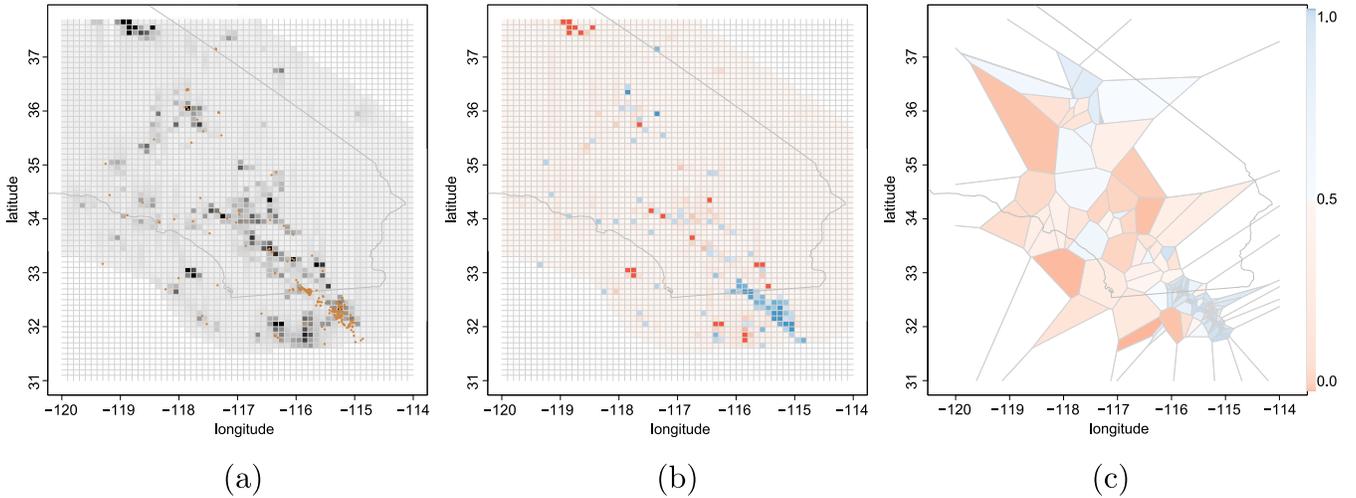}

\caption{\textup{(a)} Estimated rates under the Helmstetter, Kagan and
Jackson (\citeyear{HelKagJac07}) model, with
epicentral locations of observed earthquakes with $M \geq4.0$ in
Southern California between January 1, 2006 and January 1, 2011 overlaid.
\textup{(b)} Raw pixel residuals for Helmstetter, Kagan and
Jackson (\citeyear{HelKagJac07}) with pixels
colored according to their corresponding $p$-values.
\textup{(c)} Voronol residuals for Helmstetter, Kagan and
Jackson (\citeyear{HelKagJac07}) with pixels
colored according to their corresponding $p$-values.}
\label{HelmResPlots}
\end{figure*}

The first model under consideration is one that was submitted to RELM
by \citet{HelKagJac07} and is described in Section~\ref{sec3}. The left
panel of Figure~\ref{HelmResPlots} shows the estimated number of
earthquakes in every pixel in the greater California region that were
part of the prediction experiment. Pixels shaded very light gray have a
forecast of near zero earthquakes, while pixels shaded black forecast
much greater seismicity. The tan circles are the epicenters of the 232
earthquakes in the catalog, many of which are concentrated just South
of the Salton Sea, near the border between California and Mexico.

The extent to which the observed seismicity is in agreement with the
forecast can be visualized in the raw pixel residual plot (center panel).
The pixels are those established by the RELM experiment. Pixels where
the model predicted more events than were observed are shaded in red;
pixels where there was underprediction are shown in blue.
The degree of color saturation indicates the $p$-value of the observed
residual in the context of the forecasted Poisson distribution.
Thus, while the \citet{HelKagJac07} model greatly underpredicted
the number of events in the Salton Sea trough (dark blue), it also
forecasted a high level of seismicity in several isolated pixels that
experienced no earthquakes (dark red).
The majority of the pixels are shaded very light red, indicating
regions where the model forecast a very low rate of seismicity and no
earthquakes were recorded.


The Voronoi residual plot for the Helmstetter, Kagan and
Jackson (\citeyear{HelKagJac07}) model is
shown in the right panel of Figure~\ref{HelmResPlots}.
The spatial adaptivity of this partition is evidenced by the small
tiles in regions of high point density and larger tiles in low density
regions. The region of consistent underprediction in the Salton Sea
trough is easily identified. Unlike the raw pixel residual plot, the
Voronoi plot appears to distinguish between areas where the high
isolated rates can be considered substantial overprediction (dark red)
and areas where, considered in the context of the larger tile, the
overprediction is less extreme (light red).

\begin{figure*}

\includegraphics{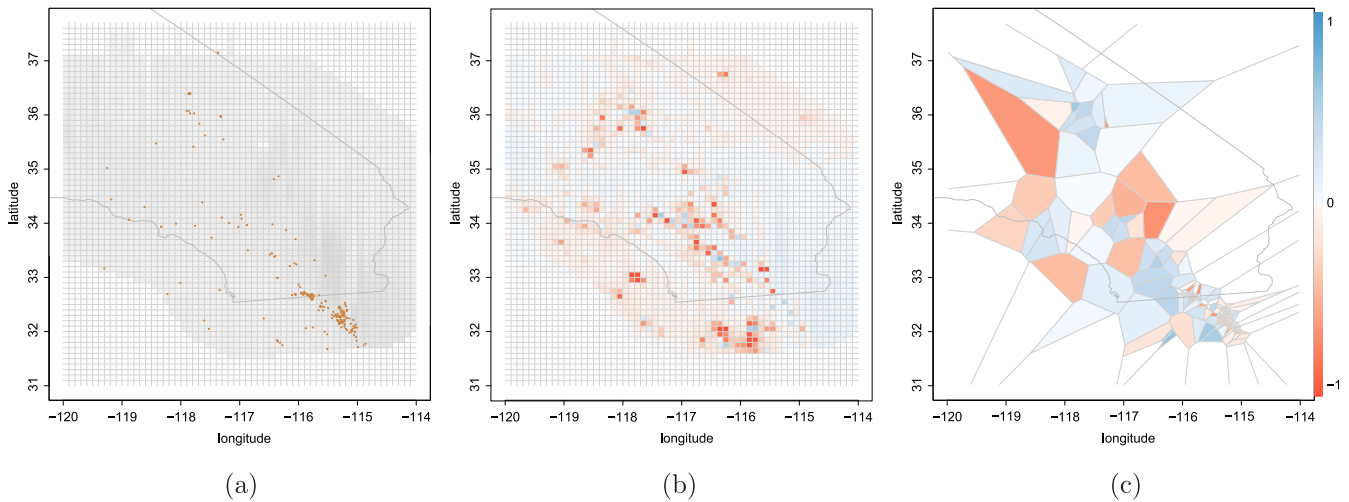}

\caption{\textup{(a)} Estimated rates under the Shen, Jackson and Kagan (\citeyear{SheJacKag07}) model, with
epicentral locations of observed earthquakes with $M \geq4.0$ in
Southern California between January 1, 2006 and January 1, 2011 overlaid.
\textup{(b)} Pixel deviance plot with blue favoring model A, Helmstetter, Kagan and
Jackson (\citeyear{HelKagJac07}),
versus model B, Shen, Jackson and Kagan (\citeyear{SheJacKag07}). Coloration is on a linear scale.
\textup{(c)} Voronoi deviance plot with blue favoring model A, Helmstetter, Kagan and
Jackson (\citeyear{HelKagJac07}), versus model B, Shen, Jackson and Kagan (\citeyear{SheJacKag07}). Coloration is on a
linear scale.}
\label{3DevPlots}
\end{figure*}

In Figure~\ref{3DevPlots} we assess how well the Helmstetter, Kagan and
Jackson (\citeyear{HelKagJac07}) model performs relative to another model in RELM using deviance
residuals. The Shen, Jackson and Kagan (\citeyear{SheJacKag07}) model is notable for utilizing
geodetic strain-rate information from past earthquakes as a proxy for
the density (intensity) of the process. $\mu(\cdot)$ is then an
interpolation of this data catalog. The result is a forecast that is
generally much smoother than the Helmstetter, Kagan and
Jackson (\citeyear{HelKagJac07}) forecast,
as seen in the left panel of Figure~\ref{3DevPlots}. The center panel
displays the deviance residuals for the Helmstetter, Kagan and
Jackson (\citeyear{HelKagJac07})
model relative to the Shen, Jackson and Kagan (\citeyear{SheJacKag07}) model.
The color scale is mapped to a measure of the comparative performance
of the two models ranging from 1 (dark blue) indicating better
performance of the Helmstetter, Kagan and
Jackson (\citeyear{HelKagJac07}) model to $-1$ (dark red)
indicating better performance of the Shen, Jackson and Kagan (\citeyear{SheJacKag07}) model.
This deviance residual plot reveals that the Helmstetter, Kagan and
Jackson (\citeyear{HelKagJac07})
model's relative advantage is in broad areas off of the main fault
lines where the forecast was lower and there were no recorded earthquakes.
It appeared to fit worse than the Shen, Jackson and Kagan (\citeyear{SheJacKag07}) model, however,
just West of the Salton Sea trough region of high seismicity, in a
swath off the coast, and in isolated pixels in central California.

The Voronoi deviance plot (right panel) identifies the same relative
underperformance of the Helmstetter, Kagan and
Jackson (\citeyear{HelKagJac07}) model relative to
the Shen, Jackson and Kagan (\citeyear{SheJacKag07}) model in the central California region and off
the coast and is a bit more informative in the areas of higher recorded
seismicity.
In the Salton Sea trough region, just south of the border of California
with Mexico, the Helmstetter, Kagan and
Jackson (\citeyear{HelKagJac07}) model appears to outperform
the Shen, Jackson and Kagan (\citeyear{SheJacKag07}) model in a vertical swath on the Western side
of the seismicity, while the results on the Eastern side are more mixed.
While these regions appear nearly white in the pixel deviance residual
plot, suggesting roughly equivalent performance of the models, the
aggregation of many of those pixels in the Voronoi plot allows for a
stronger comparison of the two models.

The utility of residual methods can be seen by contrasting the residual
plots with the error diagram of these same two models (Figure~\ref{ED}
in Section~\ref{sec5}). While the error diagram and other functional summaries
collapse the model and the observations into a new measure (such as the
false negative rate), residual methods preserve the spatial
referencing, which can help inform subsequent model generation.

\section{Discussion}\label{sec8}
The paradigm established by RELM and CSEP is a very promising direction
for earthquake model development. In addition to requiring the full
transparent specification of earthquake forecasts before the beginning
of the experiment, the criteria on which these models would be
evaluated, namely, the $L$, $N$ and $R$ tests, was also established. As
the first RELM experiment proceeded, it became apparent that these
tests can be useful summaries of the degree to which one model appears
to agree with observed seismicity, but that they leave much to be
desired. They are not well-suited to the purpose of comparing the
goodness of fit of competing models or to suggesting where models may
be improved.
It is worth noting that numerical tests such as the $L$-test, can be
viewed as examples of scoring rules (see \cite*{GneRaf07}),
and developing research on scoring rules may result in numerical tests
of improved power and efficiency.

Future prediction experiments will allow for the implementation of more
useful assessment tools. Re\-siduals methods, including super-thinned,
pixel and\break Voronoi residuals, seem ideal for comparison and to see where
a particular model appears to overpredict or underpredict seismicity.
Deviance residuals are useful for comparing two competing models and
seeing where one appears to outperform another in terms of agreement
with the observed seismicity. These methods are particularly useful in
the CSEP paradigm, as insight gained during one prediction experiment
can inform the building of models for subsequent experiments.

A note of caution should be made concerning the use of these model
assessment tools. It is common to estimate the intensity function
nonparametrically, for example using a kernel smoother. If the
selection of the tuning parameter is done while simultaneously
assessing the fit of the resulting models, this will likely lead to a
model that is overfitted. A~simple way to avoid this danger is to have
a clear separation between the model fitting stage and the model
assessment stage, as occurs when models are developed for prospective
experiments.

Although the best fitting models for forecasting earthquake occurrences
involve clustering and are thus highly non-Poissonian, it is unclear
whether the Poisson assumption implicit in the \textit{evaluation} of
these models in CSEP or RELM has anything more than a negligible impact
on the results. Since the quadrats used in these forecast evaluations
are rather large, the dependence between the numbers of events
occurring in adjacent pixels may be slight after accounting for
inhomogeneity. Further, a departure from the Poisson distribution for
the number of events occurring within a given cell would typically have
similar impacts on competing forecast models and thus have little
noticeable effect when it comes to evaluation of the relative
performance of competing models.
Nonetheless, further study is needed to clarify the importance of this
assumption in the CSEP model evaluation framework. An alternative
approach to the Poisson model would be to require that modelers provide
not only the expected number of earthquakes within each bin, but also
the joint probability distribution of counts within the bins.



Although this paper has focused on assessment tools for earthquake
models, there is a wide range of point process models to which these
methods can be applied. Super-thinned residuals and the $K$-function have
been useful in assessing models of invasive\break  species (\cite*{Betal2012}). Other recent examples, such as the use of functional summaries
in a study of infectious disease, can be found in Gelfand et al. (\citeyear{GeEtAl10}).

\section*{Acknowledgments} We thank the Editor, Associate Editor and
referees for very thoughtful remarks which substantially improved this paper.





\begin{thebibliography}{62}

\bibitem[\protect\citeauthoryear{Adelfio and Schoenberg}{2009}]{AdeSch09}
\begin{barticle}[mr]
\bauthor{\bsnm{Adelfio},~\bfnm{Giada}\binits{G.}} \AND
\bauthor{\bsnm{Schoenberg},~\bfnm{Frederic~Paik}\binits{F.~P.}}
(\byear{2009}).
\btitle{Point process diagnostics based on weighted second-order statistics and
their asymptotic properties}.
\bjournal{Ann. Inst. Statist. Math.}
\bvolume{61}
\bpages{929--948}.
\bid{doi={10.1007/s10463-008-0177-1}, issn={0020-3157}, mr={2556772}}
\bptok{imsref}%
\end{barticle}
\endbibitem

\bibitem[\protect\citeauthoryear{Baddeley, M{\o}ller and
Waagepetersen}{2000}]{BadMllWaa00}
\begin{barticle}[mr]
\bauthor{\bsnm{Baddeley},~\bfnm{A.~J.}\binits{A.~J.}},
\bauthor{\bsnm{M{\o}ller},~\bfnm{J.}\binits{J.}} \AND
\bauthor{\bsnm{Waagepetersen},~\bfnm{R.}\binits{R.}}
(\byear{2000}).
\btitle{Non- and semi-parametric estimation of interaction in inhomogeneous
point patterns}.
\bjournal{Stat. Neerl.}
\bvolume{54}
\bpages{329--350}.
\bid{doi={10.1111/1467-9574.00144}, issn={0039-0402}, mr={1804002}}
\bptok{imsref}%
\end{barticle}
\endbibitem

\bibitem[\protect\citeauthoryear{Baddeley, M{\o}ller and
Pakes}{2008}]{BadMllPak08}
\begin{barticle}[mr]
\bauthor{\bsnm{Baddeley},~\bfnm{A.}\binits{A.}},
\bauthor{\bsnm{M{\o}ller},~\bfnm{J.}\binits{J.}} \AND
\bauthor{\bsnm{Pakes},~\bfnm{A.~G.}\binits{A.~G.}}
(\byear{2008}).
\btitle{Properties of residuals for spatial point processes}.
\bjournal{Ann. Inst. Statist. Math.}
\bvolume{60}
\bpages{627--649}.
\bid{doi={10.1007/s10463-007-0116-6}, issn={0020-3157}, mr={2434415}}
\bptok{imsref}%
\end{barticle}
\endbibitem

\bibitem[\protect\citeauthoryear{Baddeley et~al.}{2005}]{Badetal05}
\begin{barticle}[mr]
\bauthor{\bsnm{Baddeley},~\bfnm{A.}\binits{A.}},
\bauthor{\bsnm{Turner},~\bfnm{R.}\binits{R.}},
\bauthor{\bsnm{M{\o}ller},~\bfnm{J.}\binits{J.}} \AND
\bauthor{\bsnm{Hazelton},~\bfnm{M.}\binits{M.}}
(\byear{2005}).
\btitle{Residual analysis for spatial point processes}.
\bjournal{J. R. Stat. Soc. Ser. B Stat. Methodol.}
\bvolume{67}
\bpages{617--666}.
\bid{doi={10.1111/j.1467-9868.2005.00519.x}, issn={1369-7412}, mr={2210685}}
\bptnote{check related}%
\bptok{imsref}%
\end{barticle}
\endbibitem

\bibitem[\protect\citeauthoryear{Bakun et~al.}{2005}]{Baketal05}
\begin{barticle}[pbm]
\bauthor{\bsnm{Bakun},~\bfnm{W.~H.}\binits{W.~H.}},
\bauthor{\bsnm{Aagaard},~\bfnm{B.}\binits{B.}},
\bauthor{\bsnm{Dost},~\bfnm{B.}\binits{B.}},
\bauthor{\bsnm{Ellsworth},~\bfnm{W.~L.}\binits{W.~L.}},
\bauthor{\bsnm{Hardebeck},~\bfnm{J.~L.}\binits{J.~L.}},
\bauthor{\bsnm{Harris},~\bfnm{R.~A.}\binits{R.~A.}},
\bauthor{\bsnm{Ji},~\bfnm{C.}\binits{C.}},
\bauthor{\bsnm{Johnston},~\bfnm{M~J~S}\binits{M.~J.~S.}},
\bauthor{\bsnm{Langbein},~\bfnm{J.}\binits{J.}},
\bauthor{\bsnm{Lienkaemper},~\bfnm{J.~J.}\binits{J.~J.}},
\bauthor{\bsnm{Michael},~\bfnm{A.~J.}\binits{A.~J.}},
\bauthor{\bsnm{Murray},~\bfnm{J.~R.}\binits{J.~R.}},
\bauthor{\bsnm{Nadeau},~\bfnm{R.~M.}\binits{R.~M.}},
\bauthor{\bsnm{Reasenberg},~\bfnm{P.~A.}\binits{P.~A.}},
\bauthor{\bsnm{Reichle},~\bfnm{M.~S.}\binits{M.~S.}},
\bauthor{\bsnm{Roeloffs},~\bfnm{E.~A.}\binits{E.~A.}},
\bauthor{\bsnm{Shakal},~\bfnm{A.}\binits{A.}},
\bauthor{\bsnm{Simpson},~\bfnm{R.~W.}\binits{R.~W.}} \AND
\bauthor{\bsnm{Waldhauser},~\bfnm{F.}\binits{F.}}
(\byear{2005}).
\btitle{Implications for prediction and hazard assessment from the 2004
Parkfield earthquake}.
\bjournal{Nature}
\bvolume{437}
\bpages{969--974}.
\bid{doi={10.1038/nature04067}, issn={1476-4687}, pii={nature04067},
pmid={16222291}}
\bptok{imsref}%
\end{barticle}
\endbibitem

\bibitem[\protect\citeauthoryear{Balderama et~al.}{2012}]{Betal2012}
\begin{barticle}[mr]
\bauthor{\bsnm{Balderama},~\bfnm{Earvin}\binits{E.}},
  \bauthor{\bsnm{Schoenberg},~\bfnm{Frederic}\binits{F.~P.}},
  \bauthor{\bsnm{Murray},~\bfnm{Erin}\binits{E.}} \AND
  \bauthor{\bsnm{Rundel},~\bfnm{Philip~W.}\binits{P.~W.}}
(\byear{2012}).
\btitle{Application of branching models in the study of invasive species}.
\bjournal{J. Amer. Statist. Assoc.}
\bvolume{107}
\bpages{467--476}.
\bid{doi={10.1080/01621459.2011.641402}, issn={0162-1459}, mr={2980058}}
\bptok{imsref}%
\end{barticle}
\endbibitem



\bibitem[\protect\citeauthoryear{Bolt}{2003}]{Bol03}
\begin{bbook}[auto:STB|2013/09/19|12:14:10]
\bauthor{\bsnm{Bolt},~\bfnm{B.}\binits{B.}}
(\byear{2003}).
\btitle{Earthquakes},
\bedition{5th} ed.
\bpublisher{Freeman}, \blocation{New York}.
\bptok{imsref}%
\end{bbook}
\endbibitem


\bibitem[\protect\citeauthoryear{Bray et~al.}{2014}]{Wonetal}
\begin{bmisc}[auto:STB|2013/09/19|12:14:10]
\bauthor{\bsnm{Bray},~\bfnm{A.}\binits{A.}},
\bauthor{\bsnm{Wong},~\bfnm{K.}\binits{K.}},
\bauthor{\bsnm{Barr},~\bfnm{C.}\binits{C.}} \AND
\bauthor{\bsnm{Schoenberg},~\bfnm{F.~P.}\binits{F.~P.}}
(\byear{2014}).
\bhowpublished{Residuals for spatial point
processes based on Voronoi tessellations. \textit{Ann. Appl. Stat.} To appear}.
\bptok{imsref}%
\end{bmisc}
\endbibitem


\bibitem[\protect\citeauthoryear{Br{\'e}maud}{1981}]{Bre81}
\begin{bbook}[mr]
\bauthor{\bsnm{Br{\'e}maud},~\bfnm{Pierre}\binits{P.}}
(\byear{1981}).
\btitle{Point Processes and Queues: Martingale Dynamics}.
\bpublisher{Springer}, \blocation{New York}.
\bid{mr={0636252}}
\bptok{imsref}%
\end{bbook}
\endbibitem

\bibitem[\protect\citeauthoryear{Chu et~al.}{2011}]{Chuetal11}
\begin{barticle}[auto:STB|2013/09/19|12:14:10]
\bauthor{\bsnm{Chu},~\bfnm{A.}\binits{A.}},
\bauthor{\bsnm{Schoenberg},~\bfnm{F.~P.}\binits{F.~P.}},
\bauthor{\bsnm{Bird},~\bfnm{P.}\binits{P.}},
\bauthor{\bsnm{Jackson},~\bfnm{D.~D.}\binits{D.~D.}} \AND
\bauthor{\bsnm{Kagan},~\bfnm{Y.~Y.}\binits{Y.~Y.}}
(\byear{2011}).
\btitle{Comparison of ETAS parameter estimates across different global tectonic
zones}.
\bjournal{Bull. Seismol. Soc. Amer.}
\bvolume{101}
\bpages{2323--2339}.
\bptok{imsref}%
\end{barticle}
\endbibitem

\bibitem[\protect\citeauthoryear{Clements, Schoenberg and
Schorlemmer}{2011}]{CleSchSch11}
\begin{barticle}[mr]
\bauthor{\bsnm{Clements},~\bfnm{Robert~Alan}\binits{R.~A.}},
\bauthor{\bsnm{Schoenberg},~\bfnm{Frederic~Paik}\binits{F.~P.}} \AND
\bauthor{\bsnm{Schorlemmer},~\bfnm{Danijel}\binits{D.}}
(\byear{2011}).
\btitle{Residual analysis methods for space--time point processes with
applications to earthquake forecast models in {C}alifornia}.
\bjournal{Ann. Appl. Stat.}
\bvolume{5}
\bpages{2549--2571}.
\bid{doi={10.1214/11-AOAS487}, issn={1932-6157}, mr={2907126}}
\bptok{imsref}%
\end{barticle}
\endbibitem

\bibitem[\protect\citeauthoryear{Clements, Schoenberg and Veen}{2012}]{CSV12}
\begin{barticle}[mr]
\bauthor{\bsnm{Clements},~\bfnm{Robert~Alan}\binits{R.~A.}},
  \bauthor{\bsnm{Schoenberg},~\bfnm{Frederic~Paik}\binits{F.~P.}} \AND
  \bauthor{\bsnm{Veen},~\bfnm{Alejandro}\binits{A.}}
(\byear{2012}).
\btitle{Evaluation of space-time point process models using super-thinning}.
\bjournal{Environmetrics}
\bvolume{23}
\bpages{606--616}.
\bid{doi={10.1002/env.2168}, issn={1180-4009}, mr={3020078}}
\bptok{imsref}%
\end{barticle}
\endbibitem

\bibitem[\protect\citeauthoryear{Console, Murru and
Falcone}{2010}]{ConMurFal10}
\begin{barticle}[auto:STB|2013/09/19|12:14:10]
\bauthor{\bsnm{Console},~\bfnm{R.}\binits{R.}},
\bauthor{\bsnm{Murru},~\bfnm{M.}\binits{M.}} \AND
\bauthor{\bsnm{Falcone},~\bfnm{G.}\binits{G.}}
(\byear{2010}).
\btitle{Probability gains of an epidemic-type aftershock sequence model in
retrospective forecasting of $M\geq  5$ earthquakes in Italy}.
\bjournal{J. Seismology}
\bvolume{14}
\bpages{9--26}.
\bptok{imsref}%
\end{barticle}
\endbibitem


\bibitem[\protect\citeauthoryear{Field}{2007}]{Fie07}
\begin{barticle}[auto:STB|2013/09/19|12:14:10]
\bauthor{\bsnm{Field},~\bfnm{E.~H.}\binits{E.~H.}}
(\byear{2007}).
\btitle{Overview of the working group for the development of regional
earthquake models (RELM)}.
\bjournal{Seismological Research Letters}
\bvolume{78}
\bpages{7--16}.
\bptok{imsref}%
\end{barticle}
\endbibitem

\bibitem[\protect\citeauthoryear{Field et~al.}{2009}]{Fieetal09}
\begin{barticle}[auto:STB|2013/09/19|12:14:10]
\bauthor{\bsnm{Field},~\bfnm{E.~H.}\binits{E.~H.}},
\bauthor{\bsnm{Dawson},~\bfnm{T.~E.}\binits{T.~E.}},
\bauthor{\bsnm{Felzer},~\bfnm{K.~R.}\binits{K.~R.}},
\bauthor{\bsnm{Frankel},~\bfnm{A.~D.}\binits{A.~D.}},
\bauthor{\bsnm{Gupta},~\bfnm{V.}\binits{V.}},
\bauthor{\bsnm{Jordan},~\bfnm{T.~H.}\binits{T.~H.}},
\bauthor{\bsnm{Parsons},~\bfnm{T.}\binits{T.}},
\bauthor{\bsnm{Petersen},~\bfnm{M.~D.}\binits{M.~D.}},
\bauthor{\bsnm{Stein},~\bfnm{R.~S.}\binits{R.~S.}},
\bauthor{\bsnm{Weldon},~\bfnm{R.~J.}\binits{R.~J.}} \AND
\bauthor{\bsnm{Wills},~\bfnm{C.~J.}\binits{C.~J.}}
(\byear{2009}).
\btitle{Uniform California Earthquake Rupture Forecast, Version 2 (UCERF 2)}.
\bjournal{Bull. Seismol. Soc. Amer.}
\bvolume{99}
\bpages{2053--2107}.
\bptok{imsref}%
\end{barticle}
\endbibitem


\bibitem[\protect\citeauthoryear{Gelfand et al.}{2010}]{GeEtAl10}
\begin{bbook}[mr]
\beditor{\bsnm{Gelfand},~\bfnm{A.}\binits{A.}},
\beditor{\bsnm{Diggle},~\bfnm{P.}\binits{P.}},
\beditor{\bsnm{Guttorp},~\bfnm{P.}\binits{P.}} \AND
\beditor{\bsnm{Fuentes},~\bfnm{M.}\binits{M.}}, eds.
(\byear{2010}).
\btitle{Handbook of Spatial Statistics}.
\bpublisher{CRC Press}, \blocation{Boca Raton, FL}.
\bid{doi={10.1201/9781420072884}, mr={2761512}}
\bptok{imsref}%
\end{bbook}
\endbibitem

\bibitem[\protect\citeauthoryear{Geller et~al.}{1997}]{Geletal97}
\begin{barticle}[auto:STB|2013/09/19|12:14:10]
\bauthor{\bsnm{Geller},~\bfnm{R.~J.}\binits{R.~J.}},
\bauthor{\bsnm{Jackson},~\bfnm{D.~D.}\binits{D.~D.}},
\bauthor{\bsnm{Kagan},~\bfnm{Y.~Y.}\binits{Y.~Y.}} \AND
\bauthor{\bsnm{Mulargia},~\bfnm{F.}\binits{F.}}
(\byear{1997}).
\btitle{Earthquakes cannot be predicted}.
\bjournal{Science}
\bvolume{275}
\bpages{1616--1617}.
\bptok{imsref}%
\end{barticle}
\endbibitem

\bibitem[\protect\citeauthoryear{Gneiting and Raftery}{2007}]{GneRaf07}
\begin{barticle}[mr]
\bauthor{\bsnm{Gneiting},~\bfnm{Tilmann}\binits{T.}} \AND
\bauthor{\bsnm{Raftery},~\bfnm{Adrian~E.}\binits{A.~E.}}
(\byear{2007}).
\btitle{Strictly proper scoring rules, prediction, and estimation}.
\bjournal{J. Amer. Statist. Assoc.}
\bvolume{102}
\bpages{359--378}.
\bid{doi={10.1198/016214506000001437}, issn={0162-1459}, mr={2345548}}
\bptok{imsref}%
\end{barticle}
\endbibitem

\bibitem[\protect\citeauthoryear{Hauksson and Goddard}{1981}]{HauGod81}
\begin{barticle}[auto:STB|2013/09/19|12:14:10]
\bauthor{\bsnm{Hauksson},~\bfnm{E.}\binits{E.}} \AND
\bauthor{\bsnm{Goddard},~\bfnm{J.~G.}\binits{J.~G.}}
(\byear{1981}).
\btitle{Radon earthquake precursor studies in Iceland}.
\bjournal{J. Geophys. Res.}
\bvolume{86}
\bpages{7037--7054}.
\bptok{imsref}%
\end{barticle}
\endbibitem

\bibitem[\protect\citeauthoryear{Hawkes}{1971}]{Haw71}
\begin{barticle}[mr]
\bauthor{\bsnm{Hawkes},~\bfnm{Alan~G.}\binits{A.~G.}}
(\byear{1971}).
\btitle{Point spectra of some mutually exciting point processes}.
\bjournal{J. R. Stat. Soc. Ser. B Stat. Methodol.}
\bvolume{33}
\bpages{438--443}.
\bid{issn={0035-9246}, mr={0358976}}
\bptok{imsref}%
\end{barticle}
\endbibitem

\bibitem[\protect\citeauthoryear{Helmstetter, Kagan and
Jackson}{2007}]{HelKagJac07}
\begin{barticle}[auto:STB|2013/09/19|12:14:10]
\bauthor{\bsnm{Helmstetter},~\bfnm{A.}\binits{A.}},
\bauthor{\bsnm{Kagan},~\bfnm{Y.~Y.}\binits{Y.~Y.}} \AND
\bauthor{\bsnm{Jackson},~\bfnm{D.~D.}\binits{D.~D.}}
(\byear{2007}).
\btitle{High-resolution time-independent grid-based forecast $M\geq 5$
earthquakes in California}.
\bjournal{Seismological Research Letters}
\bvolume{78}
\bpages{78--86}.
\bptok{imsref}%
\end{barticle}
\endbibitem

\bibitem[\protect\citeauthoryear{Helmstetter and Sornette}{2003}]{HelSor03}
\begin{barticle}[auto:STB|2013/09/19|12:14:10]
\bauthor{\bsnm{Helmstetter},~\bfnm{A.}\binits{A.}} \AND
\bauthor{\bsnm{Sornette},~\bfnm{D.}\binits{D.}}
(\byear{2003}).
\btitle{Predictability in the Epidemic-Type Aftershock Sequence model of
interacting triggered seismicity}.
\bjournal{J. Geophys. Res.}
\bvolume{108}
\bpages{2482--2499}.
\bptok{imsref}%
\end{barticle}
\endbibitem

\bibitem[\protect\citeauthoryear{Hough}{2010}]{Hou10}
\begin{bbook}[auto:STB|2013/09/19|12:14:10]
\bauthor{\bsnm{Hough},~\bfnm{S.}\binits{S.}}
(\byear{2010}).
\btitle{Predicting the Unpredictable: The Tumultuous Science of Earthquake
Prediction}.
\bpublisher{Princeton Univ. Press}, \blocation{Princeton, NJ}.
\bptok{imsref}%
\end{bbook}
\endbibitem

\bibitem[\protect\citeauthoryear{Jackson}{1996}]{Jac96}
\begin{barticle}[auto:STB|2013/09/19|12:14:10]
\bauthor{\bsnm{Jackson},~\bfnm{D.~D.}\binits{D.~D.}}
(\byear{1996}).
\btitle{Earthquake prediction evaluation standards applied to the VAN method}.
\bjournal{Geophys. Res. Lett.}
\bvolume{23}
\bpages{1363--1366}.
\bptok{imsref}%
\end{barticle}
\endbibitem

\bibitem[\protect\citeauthoryear{Jordan}{2006}]{Jor06}
\begin{barticle}[auto:STB|2013/09/19|12:14:10]
\bauthor{\bsnm{Jordan},~\bfnm{T.~H.}\binits{T.~H.}}
(\byear{2006}).
\btitle{Earthquake predictability, brick by brick}.
\bjournal{Seismological Research Letters}
\bvolume{77}
\bpages{3--6}.
\bptok{imsref}%
\end{barticle}
\endbibitem

\bibitem[\protect\citeauthoryear{Jordan and Jones}{2010}]{JorJon10}
\begin{barticle}[auto:STB|2013/09/19|12:14:10]
\bauthor{\bsnm{Jordan},~\bfnm{T.~H.}\binits{T.~H.}} \AND
\bauthor{\bsnm{Jones},~\bfnm{L.~M.}\binits{L.~M.}}
(\byear{2010}).
\btitle{Operational earthquake forecasting: Some thoughts on why and how}.
\bjournal{Seismological Research Letters}
\bvolume{81}
\bpages{571--574}.
\bptok{imsref}%
\end{barticle}
\endbibitem

\bibitem[\protect\citeauthoryear{Kagan}{1997}]{Kag97}
\begin{barticle}[auto:STB|2013/09/19|12:14:10]
\bauthor{\bsnm{Kagan},~\bfnm{Y.~Y.}\binits{Y.~Y.}}
(\byear{1997}).
\btitle{Are earthquakes predictable?}
\bjournal{Geophys. J. Int.}
\bvolume{131}
\bpages{505--525}.
\bptok{imsref}%
\end{barticle}
\endbibitem

\bibitem[\protect\citeauthoryear{Kagan}{2009}]{K09}
\begin{barticle}[auto:STB|2013/10/14|10:36:11]
\bauthor{\bsnm{Kagan},~\bfnm{Y.~Y.}\binits{Y.~Y.}}
(\byear{2009}).
\btitle{Testing long-term earthquake forecasts: Likelihood methods and error
  diagrams}.
\bjournal{Geophys. J. Int.}
\bvolume{177}
\bpages{532--542}.
\bptok{imsref}%
\end{barticle}
\endbibitem

\bibitem[\protect\citeauthoryear{Keilis-Borok and Kossobokov}{1990}]{KeiKos}
\begin{barticle}[auto:STB|2013/09/19|12:14:10]
\bauthor{\bsnm{Keilis-Borok},~\bfnm{V.}\binits{V.}} \AND
\bauthor{\bsnm{Kossobokov},~\bfnm{V.~G.}\binits{V.~G.}}
(\byear{1990}).
\btitle{Premonitory activation of earthquake flow: Algorithm M8}.
\bjournal{Physics of the Earth and Planetary Interiors}
\bvolume{6}
\bpages{73--83}.
\bptok{imsref}%
\end{barticle}
\endbibitem

\bibitem[\protect\citeauthoryear{Lewis and Shedler}{1979}]{LewShe79}
\begin{barticle}[mr]
\bauthor{\bsnm{Lewis},~\bfnm{P.~A.~W.}\binits{P.~A.~W.}} \AND
\bauthor{\bsnm{Shedler},~\bfnm{G.~S.}\binits{G.~S.}}
(\byear{1979}).
\btitle{Simulation of nonhomogeneous {P}oisson processes by thinning}.
\bjournal{Naval Res. Logist. Quart.}
\bvolume{26}
\bpages{403--413}.
\bid{doi={10.1002/nav.3800260304}, issn={0028-1441}, mr={0546120}}
\bptok{imsref}%
\end{barticle}
\endbibitem

\bibitem[\protect\citeauthoryear{Manga and Wang}{2007}]{ManWan07}
\begin{bincollection}[auto:STB|2013/09/19|12:14:10]
\bauthor{\bsnm{Manga},~\bfnm{M.}\binits{M.}} \AND
\bauthor{\bsnm{Wang},~\bfnm{C.~Y.}\binits{C.~Y.}}
(\byear{2007}).
\btitle{Earthquake hydrology}.
In \bbooktitle{Treatise on Geophysics}
(\beditor{\bfnm{G.}\binits{G.} \bsnm{Schubert}}, ed.)
\bvolume{4}
\bpages{293--320}.
\bpublisher{Elsevier}, \blocation{Amsterdam}.
\bptok{imsref}%
\end{bincollection}
\endbibitem

\bibitem[\protect\citeauthoryear{Marzocchi and Zechar}{2011}]{MarZec11}
\begin{barticle}[auto:STB|2013/09/19|12:14:10]
\bauthor{\bsnm{Marzocchi},~\bfnm{W.}\binits{W.}} \AND
\bauthor{\bsnm{Zechar},~\bfnm{J.~D.}\binits{J.~D.}}
(\byear{2011}).
\btitle{Earthquake forecasting and earthquake prediction: Different approaches
for obtaining the best model}.
\bjournal{Seismological Research Letters}
\bvolume{82}
\bpages{442--448}.
\bptok{imsref}%
\end{barticle}
\endbibitem

\bibitem[\protect\citeauthoryear{Merzbach and Nualart}{1986}]{MerNua86}
\begin{barticle}[mr]
\bauthor{\bsnm{Merzbach},~\bfnm{Ely}\binits{E.}} \AND
\bauthor{\bsnm{Nualart},~\bfnm{David}\binits{D.}}
(\byear{1986}).
\btitle{A characterization of the spatial {P}oisson process and changing time}.
\bjournal{Ann. Probab.}
\bvolume{14}
\bpages{1380--1390}.
\bid{issn={0091-1798}, mr={0866358}}
\bptok{imsref}%
\end{barticle}
\endbibitem

\bibitem[\protect\citeauthoryear{Meyer}{1971}]{Mey71}
\begin{bincollection}[mr]
\bauthor{\bsnm{Meyer},~\bfnm{P.~A.}\binits{P.~A.}}
(\byear{1971}).
\btitle{D\'emonstration simplifi\'ee d'un th\'eor\`eme de {K}night}.
In \bbooktitle{S\'eminaire de {P}robabilit\'es, {V} ({U}niv. {S}trasbourg,
Ann\'ee Universitaire 1969--1970)}
\bseries{Lecture Notes in Math.}
\bvolume{191}
\bpages{191--195}.
\bpublisher{Springer}, \blocation{Berlin}.
\bid{mr={0380972}}
\bptok{imsref}%
\end{bincollection}
\endbibitem

\bibitem[\protect\citeauthoryear{Molchan}{1991}]{M91}
\begin{barticle}[auto:STB|2013/10/14|10:36:11]
\bauthor{\bsnm{Molchan},~\bfnm{G.~M.}\binits{G.~M.}}
(\byear{1991}).
\btitle{Structure of optimal strategies in earthquake prediction}.
\bjournal{Tectonophysics}
\bvolume{193}
\bpages{267--276}.
\bptok{imsref}%
\end{barticle}
\endbibitem

\bibitem[\protect\citeauthoryear{Molchan}{2010}]{M10}
\begin{barticle}[auto:STB|2013/10/14|10:36:11]
\bauthor{\bsnm{Molchan},~\bfnm{G.}\binits{G.}}
(\byear{2010}).
\btitle{Space-time earthquake prediction: The error diagrams}.
\bjournal{Pure and Applied Geophysics}
\bvolume{167}
\bpages{907--917}.
\bptok{imsref}%
\end{barticle}
\endbibitem

\bibitem[\protect\citeauthoryear{Nair}{1990}]{Nai90}
\begin{barticle}[mr]
\bauthor{\bsnm{Nair},~\bfnm{M.~Gopalan}\binits{M.~G.}}
(\byear{1990}).
\btitle{Random space change for multiparameter point processes}.
\bjournal{Ann. Probab.}
\bvolume{18}
\bpages{1222--1231}.
\bid{issn={0091-1798}, mr={1062066}}
\bptok{imsref}%
\end{barticle}
\endbibitem

\bibitem[\protect\citeauthoryear{Ogata}{1978}]{Oga78}
\begin{barticle}[mr]
\bauthor{\bsnm{Ogata},~\bfnm{Yosihiko}\binits{Y.}}
(\byear{1978}).
\btitle{The asymptotic behaviour of maximum likelihood estimators for
stationary point processes}.
\bjournal{Ann. Inst. Statist. Math.}
\bvolume{30}
\bpages{243--261}.
\bid{doi={10.1007/BF02480216}, issn={0020-3157}, mr={0514494}}
\bptok{imsref}%
\end{barticle}
\endbibitem

\bibitem[\protect\citeauthoryear{Ogata}{1981}]{Oga81}
\begin{barticle}[auto:STB|2013/09/19|12:14:10]
\bauthor{\bsnm{Ogata},~\bfnm{Y.}\binits{Y.}}
(\byear{1981}).
\btitle{On Lewis' simulation method for point processes}.
\bjournal{IEEE Trans. Inform. Theory}
\bvolume{IT-27}
\bpages{23--31}.
\bptok{imsref}%
\end{barticle}
\endbibitem

\bibitem[\protect\citeauthoryear{Ogata}{1988}]{Oga88}
\begin{barticle}[auto:STB|2013/09/19|12:14:10]
\bauthor{\bsnm{Ogata},~\bfnm{Y.}\binits{Y.}}
(\byear{1988}).
\btitle{Statistical models for earthquake occurrences and residual analysis for
point processes}.
\bjournal{J. Amer. Statist. Assoc.}
\bvolume{83}
\bpages{9--27}.
\bptok{imsref}%
\end{barticle}
\endbibitem

\bibitem[\protect\citeauthoryear{Ogata}{1998}]{Oga98}
\begin{barticle}[auto:STB|2013/09/19|12:14:10]
\bauthor{\bsnm{Ogata},~\bfnm{Y.}\binits{Y.}}
(\byear{1998}).
\btitle{Space--time point process models for earthquake occurrences}.
\bjournal{Ann. Inst. Statist. Math.}
\bvolume{50}
\bpages{379--402}.
\bptok{imsref}%
\end{barticle}
\endbibitem

\bibitem[\protect\citeauthoryear{Ogata, Jones and Toda}{2003}]{OgaJonTod03}
\begin{barticle}[auto:STB|2013/09/19|12:14:10]
\bauthor{\bsnm{Ogata},~\bfnm{Y.}\binits{Y.}},
\bauthor{\bsnm{Jones},~\bfnm{L.~M.}\binits{L.~M.}} \AND
\bauthor{\bsnm{Toda},~\bfnm{S.}\binits{S.}}
(\byear{2003}).
\btitle{When and where the aftershock activity was depressed: Contrasting decay
patterns of the proximate large earthquakes in southern California}.
\bjournal{Journal of Geophysical Research}
\bvolume{108}\break
\bpages{2318}.
\bptok{imsref}%
\end{barticle}
\endbibitem

\bibitem[\protect\citeauthoryear{Okabe et~al.}{2000}]{Okaetal00}
\begin{bbook}[mr]
\bauthor{\bsnm{Okabe},~\bfnm{Atsuyuki}\binits{A.}},
\bauthor{\bsnm{Boots},~\bfnm{Barry}\binits{B.}},
\bauthor{\bsnm{Sugihara},~\bfnm{Kokichi}\binits{K.}} \AND
\bauthor{\bsnm{Chiu},~\bfnm{Sung~Nok}\binits{S.~N.}}
(\byear{2000}).
\btitle{Spatial Tessellations: Concepts and Applications of {V}oronoi
Diagrams},
\bedition{2nd} ed.
\bpublisher{Wiley}, \blocation{Chichester}.
\bid{mr={1770006}}
\bptok{imsref}%
\end{bbook}
\endbibitem

\bibitem[\protect\citeauthoryear{Rhoades et~al.}{2011}]{Rhoetal11}
\begin{barticle}[auto:STB|2013/09/19|12:14:10]
\bauthor{\bsnm{Rhoades},~\bfnm{D.~A.}\binits{D.~A.}},
\bauthor{\bsnm{Schorlemmer},~\bfnm{D.}\binits{D.}},
\bauthor{\bsnm{Gerstenberger},~\bfnm{M.~C.}\binits{M.~C.}},
\bauthor{\bsnm{Christophersen},~\bfnm{A.}\binits{A.}},
\bauthor{\bsnm{Zechar},~\bfnm{J.~D.}\binits{J.~D.}} \AND
\bauthor{\bsnm{Imoto},~\bfnm{M.}\binits{M.}}
(\byear{2011}).
\btitle{Efficient testing of earthquake forecasting models}.
\bjournal{Acta Geophysica}
\bvolume{59}
\bpages{728--747}.
\bptok{imsref}%
\end{barticle}
\endbibitem

\bibitem[\protect\citeauthoryear{Ripley}{1976}]{Rip76}
\begin{barticle}[mr]
\bauthor{\bsnm{Ripley},~\bfnm{B.~D.}\binits{B.~D.}}
(\byear{1976}).
\btitle{The second-order analysis of stationary point processes}.
\bjournal{J. Appl. Probab.}
\bvolume{13}
\bpages{255--266}.
\bid{issn={0021-9002}, mr={0402918}}
\bptok{imsref}%
\end{barticle}
\endbibitem

\bibitem[\protect\citeauthoryear{Schoenberg}{1999}]{Sch99}
\begin{barticle}[mr]
\bauthor{\bsnm{Schoenberg},~\bfnm{Frederic}\binits{F.}}
(\byear{1999}).
\btitle{Transforming spatial point processes into {P}oisson processes}.
\bjournal{Stochastic Process. Appl.}
\bvolume{81}
\bpages{155--164}.
\bid{doi={10.1016/S0304-4149(98)00098-2}, issn={0304-4149}, mr={1694573}}
\bptok{imsref}%
\end{barticle}
\endbibitem

\bibitem[\protect\citeauthoryear{Schoenberg}{2003}]{Sch03}
\begin{barticle}[mr]
\bauthor{\bsnm{Schoenberg},~\bfnm{Frederic~Paik}\binits{F.~P.}}
(\byear{2003}).
\btitle{Multidimensional residual analysis of point process models for
earthquake occurrences}.
\bjournal{J. Amer. Statist. Assoc.}
\bvolume{98}
\bpages{789--795}.
\bid{doi={10.1198/016214503000000710}, issn={0162-1459}, mr={2055487}}
\bptok{imsref}%
\end{barticle}
\endbibitem

\bibitem[\protect\citeauthoryear{Schoenberg}{2013}]{Sch13}
\begin{barticle}[auto:STB|2013/09/19|12:14:10]
\bauthor{\bsnm{Schoenberg},~\bfnm{F.~P.}\binits{F.~P.}}
(\byear{2013}).
\btitle{Facilitated estimation of ETAS}.
\bjournal{Bulletin of the Seismological Society of America}
\bvolume{103}
\bpages{1--7}.
\bptok{imsref}%
\end{barticle}
\endbibitem


\bibitem[\protect\citeauthoryear{Schorlemmer et~al.}{2007}]{Schetal07}
\begin{barticle}[auto:STB|2013/09/19|12:14:10]
\bauthor{\bsnm{Schorlemmer},~\bfnm{D.}\binits{D.}},
\bauthor{\bsnm{Gerstenberger},~\bfnm{M.~C.}\binits{M.~C.}},
\bauthor{\bsnm{Wiemer},~\bfnm{S.}\binits{S.}},
\bauthor{\bsnm{Jackson},~\bfnm{D.~D.}\binits{D.~D.}} \AND
\bauthor{\bsnm{Rhoades},~\bfnm{D.~A.}\binits{D.~A.}}
(\byear{2007}).
\btitle{Earthquake likelihood model testing}.
\bjournal{Seismological Research Letters}
\bvolume{78}
\bpages{17--27}.
\bptok{imsref}%
\end{barticle}
\endbibitem


\bibitem[\protect\citeauthoryear{Shen, Jackson and Kagan}{2007}]{SheJacKag07}
\begin{barticle}[auto:STB|2013/09/19|12:14:10]
\bauthor{\bsnm{Shen},~\bfnm{Z.~K.}\binits{Z.~K.}},
\bauthor{\bsnm{Jackson},~\bfnm{D.~D.}\binits{D.~D.}} \AND
\bauthor{\bsnm{Kagan},~\bfnm{Y.~Y.}\binits{Y.~Y.}}
(\byear{2007}).
\btitle{Implications of geodetic strain rate for future earthquakes, with a
five-year forecast of M5 earthquakes in southern California}.
\bjournal{Seismological Research Letters}
\bvolume{78}
\bpages{116--120}.
\bptok{imsref}%
\end{barticle}
\endbibitem

\bibitem[\protect\citeauthoryear{Sornette}{2005}]{Sor05}
\begin{barticle}[auto:STB|2013/09/19|12:14:10]
\bauthor{\bsnm{Sornette},~\bfnm{D.}\binits{D.}}
(\byear{2005}).
\btitle{Apparent clustering and apparent background earthquakes biased by
undetected seismicity}.
\bjournal{J. Geophys. Res.}
\bvolume{110}
\bpages{B09303}.
\bptok{imsref}%
\end{barticle}
\endbibitem

\bibitem[\protect\citeauthoryear{Swets}{1973}]{S73}
\begin{barticle}[pbm]
\bauthor{\bsnm{Swets},~\bfnm{J.~A.}\binits{J.~A.}}
(\byear{1973}).
\btitle{The relative operating characteristic in psychology}.
\bjournal{Science}
\bvolume{182}
\bpages{990--1000}.
\bid{doi={10.1126/science.182.4116.990}, issn={0036-8075}, pii={182/4116/990},
  pmid={17833780}}
\bptok{imsref}%
\end{barticle}
\endbibitem

\bibitem[\protect\citeauthoryear{Tiampo and Shcherbakov}{2012}]{TiaShc12}
\begin{barticle}[auto:STB|2013/09/19|12:14:10]
\bauthor{\bsnm{Tiampo},~\bfnm{K.~R.}\binits{K.~R.}} \AND
\bauthor{\bsnm{Shcherbakov},~\bfnm{R.}\binits{R.}}
(\byear{2012}).
\btitle{Seismicity-based earthquake forecasting techniques: Ten years of
progress}.
\bjournal{Tectonophysics}
\bvolume{522}
\bpages{89--121}.
\bptok{imsref}%
\end{barticle}
\endbibitem

\bibitem[\protect\citeauthoryear{Veen and Schoenberg}{2006}]{VeeSch06}
\begin{bincollection}[mr]
\bauthor{\bsnm{Veen},~\bfnm{Alejandro}\binits{A.}} \AND
\bauthor{\bsnm{Schoenberg},~\bfnm{Frederic~Paik}\binits{F.~P.}}
(\byear{2006}).
\btitle{Assessing spatial point process models using weighted {$K$}-functions:
Analysis of {C}alifornia earthquakes}.
In \bbooktitle{Case Studies in Spatial Point Process Modeling}
(\beditor{\binits{A.}\bfnm{A.} \bsnm{Baddeley}},
\beditor{\binits{P.}\bfnm{P.} \bsnm{Gregori}},
\beditor{\binits{J.}\bfnm{J.} \bsnm{Mateu}},
\beditor{\binits{R.}\bfnm{R.} \bsnm{Stoica}}
\AND
\beditor{\binits{D.}\bfnm{D.} \bsnm{Stoyan}}, eds.).
\bseries{Lecture Notes in Statist.}
\bvolume{185}
\bpages{293--306}.
\bpublisher{Springer}, \blocation{New York}.
\bid{doi={10.1007/0-387-31144-0_16}, mr={2232135}}
\bptnote{check year}%
\bptok{imsref}%
\end{bincollection}
\endbibitem

\bibitem[\protect\citeauthoryear{Vere-Jones and Schoenberg}{2004}]{VerSch04}
\begin{barticle}[mr]
\bauthor{\bsnm{Vere-Jones},~\bfnm{David}\binits{D.}} \AND
\bauthor{\bsnm{Schoenberg},~\bfnm{Frederic~Paik}\binits{F.~P.}}
(\byear{2004}).
\btitle{Rescaling marked point processes}.
\bjournal{Aust. N. Z. J. Stat.}
\bvolume{46}
\bpages{133--143}.
\bid{doi={10.1111/j.1467-842X.2004.00319.x}, issn={1369-1473}, mr={2055791}}
\bptok{imsref}%
\end{barticle}
\endbibitem

\bibitem[\protect\citeauthoryear{Vere-Jones and Zhuang}{2008}]{VerZhu08}
\begin{barticle}[auto:STB|2013/09/19|12:14:10]
\bauthor{\bsnm{Vere-Jones},~\bfnm{D.}\binits{D.}} \AND
\bauthor{\bsnm{Zhuang},~\bfnm{J.}\binits{J.}}
(\byear{2008}).
\btitle{On the distribution of the largest event in the critical ETAS model}.
\bjournal{Phys. Rev. E (3)}
\bvolume{78}
\bpages{047102}.
\bptok{imsref}%
\end{barticle}
\endbibitem

\bibitem[\protect\citeauthoryear{Wang, Jackson and Kagan}{2011}]{WanJacKag11}
\begin{barticle}[auto:STB|2013/09/19|12:14:10]
\bauthor{\bsnm{Wang},~\bfnm{Q.}\binits{Q.}},
\bauthor{\bsnm{Jackson},~\bfnm{D.~D.}\binits{D.~D.}} \AND
\bauthor{\bsnm{Kagan},~\bfnm{Y.~Y.}\binits{Y.~Y.}}
(\byear{2011}).
\btitle{California earthquake forecasts based on smoothed seismicity: Model
choices}.
\bjournal{Bull. Seismol. Soc. Amer.}
\bvolume{101}
\bpages{1422--\break1430}.
\bptok{imsref}%
\end{barticle}
\endbibitem

\bibitem[\protect\citeauthoryear{Werner et~al.}{2011}]{Weretal11}
\begin{barticle}[auto:STB|2013/09/19|12:14:10]
\bauthor{\bsnm{Werner},~\bfnm{M.~J.}\binits{M.~J.}},
\bauthor{\bsnm{Helmstetter},~\bfnm{A.}\binits{A.}},
\bauthor{\bsnm{Jackson},~\bfnm{D.~D.}\binits{D.~D.}} \AND
\bauthor{\bsnm{Kagan},~\bfnm{Y.~Y.}\binits{Y.~Y.}}
(\byear{2011}).
\btitle{High-Resolution Long-Term and Short-Term Earthquake Forecasts for
California}.
\bjournal{Bull. Seismol. Soc. Amer.}
\bvolume{101}
\bpages{1630--1648}.
\bptok{imsref}%
\end{barticle}
\endbibitem


\bibitem[\protect\citeauthoryear{Zaliapin and Molchan}{2004}]{ZM04}
\begin{barticle}[auto:STB|2013/10/14|10:36:11]
\bauthor{\bsnm{Zaliapin},~\bfnm{I.}\binits{I.}} \AND
  \bauthor{\bsnm{Molchan},~\bfnm{G.}\binits{G.}}
(\byear{2004}).
\btitle{Tossing the earth: How to reliably test earthquake prediction methods.}
\bjournal{Eos. Trans. AGU}
\bvolume{85}
\bpages{47}.
\bnote{S23A--0302}.
\bptok{imsref}%
\end{barticle}
\endbibitem



\bibitem[\protect\citeauthoryear{Zechar, Gerstenberger and
Rhoades}{2010}]{ZecGerRho10}
\begin{barticle}[auto:STB|2013/09/19|12:14:10]
\bauthor{\bsnm{Zechar},~\bfnm{J.~D.}\binits{J.~D.}},
\bauthor{\bsnm{Gerstenberger},~\bfnm{M.~C.}\binits{M.~C.}} \AND
\bauthor{\bsnm{Rhoades},~\bfnm{D.~A.}\binits{D.~A.}}
(\byear{2010}).
\btitle{Likelihood-based tests for evaluating space--rate-magnitude earthquake
forecasts}.
\bjournal{Bull. Seismol. Soc. Amer.}
\bvolume{100}
\bpages{1184--1195}.
\bptok{imsref}%
\end{barticle}
\endbibitem

\bibitem[\protect\citeauthoryear{Zechar and Jordan}{2008}]{ZecJor08}
\begin{barticle}[auto:STB|2013/09/19|12:14:10]
\bauthor{\bsnm{Zechar},~\bfnm{J.~D.}\binits{J.~D.}} \AND
\bauthor{\bsnm{Jordan},~\bfnm{T.~H.}\binits{T.~H.}}
(\byear{2008}).
\btitle{Testing alarm-based earthquake predictions}.
\bjournal{Geophys. J. Int.}
\bvolume{172}
\bpages{715--724}.
\bptok{imsref}%
\end{barticle}
\endbibitem

\bibitem[\protect\citeauthoryear{Zechar et~al.}{2013}]{Zecetal}
\begin{bmisc}[auto:STB|2013/09/19|12:14:10]
\bauthor{\bsnm{Zechar},~\bfnm{J.~D.}\binits{J.~D.}},
\bauthor{\bsnm{Schorlemmer},~\bfnm{D.}\binits{D.}},
\bauthor{\bsnm{Werner},~\bfnm{M.~J.}\binits{M.~J.}},
\bauthor{\bsnm{Gerstenberger},~\bfnm{M.~C.}\binits{M.~C.}},
\bauthor{\bsnm{Rhoades},~\bfnm{D.~A.}\binits{D.~A.}} \AND
\bauthor{\bsnm{Jordan},~\bfnm{T.~H.}\binits{T.~H.}}
(\byear{2013}).
\bhowpublished{Regional earthquake likelihood models I: First-order
results. Unpublished manuscript}.
\bptok{imsref}%
\end{bmisc}
\endbibitem

\bibitem[\protect\citeauthoryear{Zhuang}{2011}]{Zhu11}
\begin{barticle}[auto:STB|2013/09/19|12:14:10]
\bauthor{\bsnm{Zhuang},~\bfnm{J.}\binits{J.}}
(\byear{2011}).
\btitle{Next-day earthquake forecasts for the Japan region generated by the
ETAS model}.
\bjournal{Earth Planets Space}
\bvolume{63}
\bpages{207--216}.
\bptok{imsref}%
\end{barticle}
\endbibitem

\end{thebibliography}
\end{document}